\documentclass[a4paper, 10pt, twocolumn]{article}

\usepackage{amsfonts,amssymb,amsmath,amsthm}
\usepackage{subfigure}
\usepackage{graphicx}
\usepackage[footnotesize]{caption}
\usepackage{xcolor}

\usepackage[top=1.5cm, left=1.5cm, right=1.5cm, bottom=1.5cm]{geometry}

\newtheorem{proposition}{Proposition}

\def\diag{\mathbf{diag}}
\def\Diag{\mathbf{Diag}}
\def\Ker{\mathrm{Ker}}

\def\fpsf{f_{\mathrm{PSF}}}
\def\fspec{f_{\mathrm{spec}}}
\def\Dpsf{\ensuremath{\mathcal{D}_{\mathrm{PSF}}}}
\def\Dsr{\ensuremath{\mathcal{D}_{\mathrm{SR}}}}

\def\CCb{\Hb}

\input alphabet

\def\barr{\begin{array}}                \def\earr{\end{array}}
\def\bcc{\begin{center}}                \def\ecc{\end{center}}
\def\cl#1{\centerline{#1}}
\def\bdoc{\begin{document}}             \def\edoc{\end{document}}
\def\ben{\begin{enumerate}}             \def\een{\end{enumerate}}
\def\beqn{\begin{eqnarray}}             \def\eeqn{\end{eqnarray}}
\def\beqnl#1{\beqn\label{#1}}           \def\eeqnl#1{\label{#1}\eeqn}
\def\beqnx{\begin{eqnarray*}}           \def\eeqnx{\end{eqnarray*}}
\def\bseqn{\begin{subeqnarray}}         \def\eseqn{\end{subeqnarray}}
\def\beq#1\eeq{\begin{equation}#1\end{equation}}
\def\bal#1\eal{\begin{align}#1\end{align}}
\def\balx#1\ealx{\begin{align*}#1\end{align*}}
\def\beqx{$$}                           \def\eeqx{$$}
\def\bfig{\protect\begin{figure}}       \def\efig{\protect\end{figure}}
\def\bfigx{\protect\begin{figure*}}     \def\efigx{\protect\end{figure*}}
\def\bfigt{\protect\begin{figurette}}   \def\efigt{\protect\end{figurette}}
\def\bfl{\begin{flushleft}}             \def\efl{\end{flushleft}}
\def\bfr{\begin{flushright}}            \def\efr{\end{flushright}}
\def\bit{\begin{itemize}}               \def\eit{\end{itemize}}
\def\bmi{\begin{minipage}}              \def\emi{\end{minipage}}
\newcommand{\bfmi}{\begin{fminipage}}   \newcommand{\efmi}{\end{fminipage}}
\def\bpic{\begin{picture}}              \def\epic{\end{picture}}
\def\btabl{\begin{table}}               \def\etabl{\end{table}}
\def\btab{\begin{tabular}} 
\def\btabu{\begin{tabular}}             \def\etabu{\end{tabular}}
\def\btabx{\begin{tabular*}}            \def\etabx{\end{tabular*}}

\input jidef
\def\Card{\#}
\def\dd{\,d}
\def\leq{\le}
\def\geq{\ge}
\renewcommand{\det}[1]{\bars{#1}}
\def\rang{{\mathrm{rank}}}
\def\V{.}              
\def\e#1{\rm{e}^{#1}}  
\def\numero{n\ensuremath{^{\circ}}}

\def\et{{\normalfont and }}
\def\editorname{(ed.)}
\def\editornames{(eds.)}
\def\eqname{Eq.~}
\def\eqnames{Eqs~}
\def\tabname{Table~}
\def\tabnames{Tables~}
\def\figname{Figure~}
\def\fignames{Figures~}
\def\chapname{Chapter~}
\def\chapnames{Chapters~}
\def\parname{\S~}
\def\parnames{\S~}
\def\sectname{Section~}
\def\sectnames{Sections~}
\def\theoname{Theorem~}

\def\gui#1{``#1''}
\def\<{``}\def\>{''}
\def\ier{st\XS}
\def\iere{st\XS}
\def\ieme{th\XS}

\def\SI{\text{if\:}}       \def\Si{\text{If\:}}
\def\ET{\text{and\:}}       \def\OU{\text{or\:}}
\def\ALORS{\text{then\:}}
\def\DOU{\text{hence\:}}  \def\Ou{\text{where\:}}
\def\QUAND{\text{when\:}}
\def\POUR{\text{for\:}}   \def\POURTOUT{\text{for all\:}}
\def\SC{\text{s.\,t.\:}}
\def\SINON{\text{otherwise}}
\def\AVEC{\text{with\:}}
\def\DANS{\text{in\:}}
\def\IFF{\textit{if and only if}\XS}
\def\ssi{\textit{si et seulement si}\XS}

\def\post{\textit{posterior}\XS}
\def\Post{\textit{Posterior}\XS}
\def\etcoll{\textit{et al.}}
\def\andname{and\xspace}
\def\Toeplitz{Toeplitz\XS}
\def\wrt{w.r.t.\ }
\def\etal{{\itshape et al.}}
\def\etcoll{{\itshape et al.}}

\def\croext#1{\left(#1\right)}             \def\stdcroext#1{(#1)}
\def\bigcroext#1{\bigl(#1\bigr)}           \def\biggcroext#1{\biggl(#1\biggr)}
\def\Bigcroext#1{\Bigl(#1\Bigr)}           \def\Biggcroext#1{\Biggl(#1\Biggr)}

\def\TF{\text{FT}\XS} \def\TFD{\foo{DFT}} \def\TFDD{\text{DFT}\XS}
\def\FT{\text{FT}\XS} \def\DFT{\foo{DFT}} \def\DFTD{\text{DFT}\XS}

\title{Uniqueness of the Random Illumination Microscopy Variance Equation}

\author{Simon~Labouesse$^1$, J\'er\^ome~Idier$^2$, Anne~Sentenac$^1$ and Thomas Mangeat$^{3}$.\\[-1mm]
\footnotesize $^1$Aix Marseille University, CNRS, Centrale Marseille, Institut Fresnel, F-13013 Marseille, France.\\[-1mm]
\footnotesize $^2$Laboratoire des Sciences du Num\'erique de Nantes, Centrale Nantes, F-44321 Nantes, France.\\[-1mm]
\footnotesize $^3$Centre for Integrative Biology, Universit\'e de Toulouse, CNRS, UPS, F-31062 Toulouse.}	
\date{\empty} 

\renewenvironment{abstract}{\bf\small {\em\ Abstract---}}{}

\begin{document}

\maketitle

\begin{abstract}
Recently, it has been shown theoretically that fluorescence microscopy using random illuminations (RIM) yields a doubled lateral resolution and an improved optical sectionning.
Moreover, an algorithm called algoRIM, based on variance matching, has been successfully validated on numerous biological applications. Here, we propose a proof of uniqueness of the RIM variance equation, which corresponds to a first theoretical validation of algoRIM.
\end{abstract}


\section{Introduction}
\label{sec:introduction}

Fluorescence microscopy is inherently limited by diffraction to a resolution of $\approx 200$nm. Structured illumination microscopy (SIM) breaks this resolution limit by exciting the fluorophores inside the object to recover with multiple harmonic illuminations \cite{Heintzmann99,Gustafsson00}. From multiple SIM images and a precise knowledge of the illuminations, the resolution of an epifluorescence microscope can be doubled. However, the effective super-resolution (SR) capacity of SIM is often hampered by grid distorsions due to ligth scattering and optical aberrations. 

Random illumination microscopy (RIM) is an evolution of SIM which is based on speckle illuminations. Whereas SIM relies on the tight control of the illumination grids, RIM only relies on the knowledge of the speckle statistics \cite{Idier18,Mangeat20}, with a potential of increased robustness compared to SIM. 
In \cite{Idier18}, it is shown that the theoretical SR capacity of RIM is identical to that of SIM. However, the latter study exploits the statistical \emph{covariance matrix} of the recorded images, which is not a realistic scheme in terms of storage and of computing operations.

Here, we demonstrate that the knowledge of the statistical variance is actually sufficient to recover an image of the biological sample with the same SR factor as covariance-based RIM and SIM.
Our novel result concerning variance-based RIM is Proposition~\ref{unicity} in Section~\ref{sec:Variance-based RIM}, and the proof is postponed in Section~\ref{sec:proof}.


\section{Imaging model}
\label{sec:model}
For the sake of clarity, we mainly restrict ourselves to the case of 2D biological samples, and we formulate the problem in a fully discrete setting, where both the recorded images and the biological sample are represented on fine grids, with a sampling rate common to both. We assume a linear invariant response of the microscope to the incoherent fluorescence light and a linear response of the fluorophores to the intensity of the excitation coherent light. 
RIM images can then be modeled by:
\begin{equation*}
\zb_m = \yb_m + \epsilonb_m,
\end{equation*}
with
\begin{equation*}
\yb_m = \Hb \left(\rhob \circ \Ib_m \right), 
\end{equation*}
where $\epsilonb_m$ is a random variable modeling an additive noise, $\yb_m$ is a vectorized image corresponding to the $m$th illumination $\Ib_m$, $\Hb$ a convolution matrix corresponding to a convolution by the PSF $\hb$ of the microscope, $\rhob$ the fluorescence density map to recover, and $\circ$ the Hadamard matrix product \cite[Chapter\,5]{Horn91}.

The speckle covariance $\mathbf{Cov}(\Ib_m) = \Cb$ as well as the noise covariance $\mathbf{Cov}(\epsilonb_m) = \Gammab_\epsilonb$ are supposed to be known statistics.
The covariance matrix of each $\zb_m$ reads
\begin{equation*}
\Gammab_\zb(\rhob) = \Gammab_\yb(\rhob) + \Gammab_\epsilonb,
\end{equation*}
with
\begin{equation*}
\Gammab_\yb(\rhob) = \Hb \Diag(\rhob) \Cb \Diag(\rhob) \Hb\T.
\end{equation*}
The variance identifies with the diagonal of the covariance matrix $\vb_\zb(\rhob) = \diag(\Gammab_\zb)$.
The noise covariance function $\Gammab_\epsilon$ is assumed to be known. The knowledge of $\vb_\zb$ is thus equivalent to that of
\begin{equation}
    \vb_\yb = \diag(\Gammab_\yb).
    \label{eq:variance}
\end{equation} 
Hereafter, we refer to $\vb_\yb$ and $\Gammab_\yb$ as $\vb$ and $\Gammab$, respectively. 
In this document, we adopt the standard assumption of a perfect circular lens. For 2D imaging at the focal plane, the PSF \hb is a discretized Airy pattern~\cite[Sec.\,4.4.2]{Goodman96}, and the optical 
transfer function (OTF) $\wt\hb$ has a frequency cut-off $2\froc{\text{NA}}{\lambda}$, with NA the numerical aperture of the microscope and $\lambda$ the emission/excitation 
wavelength. We further assume that the illumination of the sample 
and the collection of the emitted light is performed through the same 
optical device. Ignoring the Stokes-shift, we will assume that $\Hb=\Hb\T=\Cb$.

Since our goal is to demonstrate a factor two in terms of SR, the sampling rate of the object must be at least four times the cutoff frequency imposed by the PSF. In the rest of this document, we make use of the following notations:
$\fpsf\leq1/4$ denotes the normalized cutoff frequency imposed by the PSF, and
$$
\Gc = \bigacc{\nub\in\eR^d, \norm{\nub}_\infty<1/2}\cup\bigacc{\nb/N, \nb\in\eZ^d}
$$
denotes the $d$-dimensional normalized frequency grid limited by the Nyquist frequency ($d=2$ for 2D imaging). Here, we assume that RIM acquisitions $\zb_m$ are made of $N=n^d$ elements. Then each of them can be decomposed over the set of discrete frequencies $\Dpsf=\Dc(\fpsf)$, where $\Dc(f)$ is a generic notation for the ``discrete interior'' of a ball of radius~$f$:
$$
\Dc(f)=\stdacc{\nub\in\Gc, \norm{\nub}<f}.
$$
%
%
%
%
\section{Variance-based RIM}
\subsection{SR from variance equations}
\label{sec:Variance-based RIM}
In the 2D case, \cite{Idier18} obtains that the knowledge of \Gammab allows to retrieve the frequency components of \rhob within the ball $\Dsr=\Dc(2\fspec)$, provided that the speckle illuminations have a cut-off frequency not larger than that of the PSF, \ie $\fspec\leq\fpsf$. When  $\fspec=\fpsf$, we have $\Dsr=\Dc(2\fpsf)$, which exactly corresponds to an SR factor equal to two.
\begin{proposition}
\label{unicity0}
Let \rhob be any entrywise nonnegative vector of size $N$. For any entrywise nonnegative solution \qb to the quadratic system $\Gammab(\qb)=\Gammab(\rhob)$, the frequency components of \qb coincide with that of \rhob in $\Dsr$.
\end{proposition}

The quadratic system of Proposition~\ref{unicity0} is made of $\frac12 N(N+1)$ real equations, for only $M$ free real-valued variables, where $M$ stand for the cardinality of $\Dsr$. Since $M\leq\frac{\pi}4N$ in 2D (and $M\leq N$ is 1D),
there is room left for a refined identifiability result, using a smaller number of equations. In this vein, Proposition~\ref{unicity} states that the $N$ variance equations are sufficient to uniquely determine the $M$ frequency components in $\Dsr$, provided that \rhob is an entrywise positive vector.
\begin{proposition}
\label{unicity}
Let \rhob be any entrywise positive vector of size $N$. For any entrywise nonnegative solution \qb to the quadratic system of $N$ equations $\vb(\qb)=\vb(\rhob)$, the frequency components of \qb coincide with that of \rhob in $\Dsr$, while the frequency components of \qb outside $\Dsr$ remain arbitrary (up to the nonnegativity constraint on the entries of \qb).
\end{proposition}

\subsection{Proof of Proposition~\ref{unicity}}
\label{sec:proof}
Let us define the bilinear vector-valued function:
\begin{equation}
  \label{eq:F}
\fb(\xb, \yb) = \diag\bigpth{\Hb \Diag(\xb)\CCb \Diag(\yb) \Hb},
\end{equation}
so that $\vb(\rhob)=\fb(\rhob,\rhob)$.
Each component of \fb is a symmetric form, since
$\fb(\xb, \yb) = \fb(\yb, \xb)$.
Let us define
\begin{align}
\Mb_\xb &= \Hb \Diag(\xb) \CCb,\notag\\
\Bb_\xb &= \Hb\circ\Mb_\xb,\notag
\end{align}
so that
\begin{equation}
\fb(\xb, \yb) = \Bb_\xb \yb = \Bb_\yb \xb
\label{eq:FmatrixProd}
\end{equation}
according to the matrix identity~\cite{Petersen12}
$$
\diag\pth{\Av\Diag(\vb)\Bv\T}=(\Av\circ\Bv)\,\vb=(\Bv\circ\Av)\,\vb.
$$
In particular, for a given object \rhob, the (noiseless) data variance vector \eqref{eq:variance} is given by $\fb(\rhob,\rhob) = \Bb_\rhob \rhob$.

\begin{proposition}
\label{prop:ker}
For any two real solutions \rhob and \qb to Eq.~\eqref{eq:variance}, we have $\rhob-\qb\in\Ker(\Bb_{\rhob+\qb})$ and $\rhob+\qb\in\Ker(\Bb_{\rhob-\qb})$.
\end{proposition}
\begin{proof}
Indeed,
\begin{align}
	\fb(\rhob+\qb, \rhob-\qb) &= \fb(\rhob, \rhob)  - \fb(\qb, \qb)  + \fb(\qb, \rhob)  - \fb(\rhob, \qb)  \notag\\ 
  	 &= \vb  - \vb  + \fb(\qb, \rhob)  - \fb(\qb, \rhob) \notag\\
	 &= \zerob
  	   \label{eq:Feq0}
\end{align}
Combining Equations \eqref{eq:FmatrixProd} and \eqref{eq:Feq0}, we obtain
$$
\Bb_{\rhob+\qb} (\rhob - \qb) = \Bb_{\rhob-\qb} (\rhob+\qb) = \zerob,
$$
which proves the assertion.
\end{proof}

\begin{proposition}
\label{prop:kerbis}
For any vector \xb with positive entries, $\Ker(\Bb_{\xb})$ is the linear span of frequency components outside $\Dsr$.
\end{proposition}
\begin{proof}
Let $K_{\min} = \min(\xb)$, so that $\xb_{\min} = \xb - K_{\min}$ is entrywise nonnegative.
We have $\Bb_\xb=K_{\min}\,\Gb+\Bb_{\xb_{\min}}$, with $\Gb=\Hb^2\circ\Hb$. Matrix \Gb is circulant. It can be seen as a convolution matrix with a filter $\gb=(\hb\star\hb)\circ\hb$, with $\wt\gb=(\wt\hb\circ\wt\hb)\star\wt\hb$ and $\star$ the discrete convolution. Vector $\wt\gb$ has nonzero components for all spatial frequencies belonging to $\Dsr$.
Moreover, matrix \Gb is obviously nonnegative definite.
Matrix $\Bb_{\xb_{\min}}$ is also nonnegative definite according to the Schur product Theorem, as the Hadamard product between two nonnegative definite matrices \cite[Theorem\,5.2.1]{Horn91}. 
Therefore, we have $\Ker(\Bb_\xb)=\Ker(\Gb)\cap\Ker(\Bb_{\xb_{\min}}) \subset \Ker(\Gb)$.

Similarly, let $K_{\max} = \max(\xb)$, so that $\xb_{\max} = K_{\max}-\xb$ is entrywise nonnegative.
We have $\Bb_\xb = K_{\max}\,\Gb-\Bb_{\xb_{\max}}$, and $\Bb_{\xb_{\max}}$ and $\Bb_\xb$ are both nonnegative definite. For all $\zb\in\Ker(\Gb)$, $\zb^\dag\Bb_\xb\zb=-\zb^\dag\Bb_{\xb_{\max}}\zb$, where the lhs and the rhs are nonnegative and nonpositive, respectively. We conclude that $\zb^\dag\Bb_\xb\zb=0$, so $\Ker(\Gb)\subset\Ker(\Bb_\xb)$, and finally that $\Ker(\Bb_\xb)=\Ker(\Gb)$.
\end{proof}

According to Proposition~\ref{prop:ker}, we have $\rhob-\qb\in\Ker(\Bb_{\rhob+\qb})$, where $\rhob+\qb$ is entrywise positive. Therefore, we can conclude that $\rhob-\qb\in\Ker(\Gb)$, \ie that the frequency components of \qb coincide with that of \rhob in $\Dsr$. Moreover, the frequency components of $\qb$ outside $\Dsr$ have no impact on the data covariance, and hence on its diagonal \cite{Idier18}.


\section{Conclusion}
\label{sec:conclusion}

This paper provides a mathematical proof that the super-resolution capacity of random illumination microscopy still holds when only the statistical variance of collected images is considered instead of the full covariance. Such a theoretical result meets practical evidences recently obtained concerning 2D variance-based imaging applied to various types of biological samples \cite{Mangeat20}.

Several comments can be made about the novel variance-based result, compared to its covariance-based counterpart:
\bit
\item Whereas the covariance-based result of Proposition \ref{unicity0} holds if $\fpsf=\fspec$, the proof of Proposition~\ref{unicity} is based on the fact that matrices \Hb and \Cb identify, which is more stringent. Indeed, we have a small size counter-example proving that  Proposition~\ref{unicity} is no more valid when $\Hb\neq\Cb$, even if $\fpsf=\fspec$. 
\item Another difference concerns the fact that strict positivity of the sample is needed in Proposition~\ref{unicity}. However, we have strong elements showing that this condition could be relaxed. 
\item Although we have restricted ourselves to the 2D case, a formal extension to 3D is straightforward, with the benefit of an axial super-resolution effect, on top of the lateral one obtained in 2D. For each depth, several speckle illuminations must be recorded, so that a 3D map of variance can be constructed, and a 3D map of fluorescence can be retrieved on this basis. 
\eit 

\pagebreak
\bibliographystyle{abbrv}
\bibliography{main}
\end{document}